\documentclass{emulateapj}
\newcommand{\ergss}{ergs~s$^{-1}$}

\newcommand{\chandra}{{\sl Chandra}}
\newcommand{\xmm}{{\sl XMM-Newton}}
\newcommand{\vla}{{\sl VLA}}
\newcommand{\hst}{{\sl HST}}
\newcommand{\msun}{M$_\odot$}

\newcommand{\degree}{$^\circ$}

\slugcomment{Draft version of \today}
                                
\begin{document}

\title{Radio Emission Associated with the ULX in Holmberg~II}

\author{Neal A. Miller\altaffilmark{1,2}}

\author{Richard F. Mushotzky\altaffilmark{3}}

\author{Susan G. Neff\altaffilmark{4}}

\altaffiltext{1}{Jansky Fellow of the National Radio Astronomy Observatory. The National Radio Astronomy Observatory is a facility of the National Science Foundation operated under cooperative agreement by Associated Universities, Inc.}
\altaffiltext{2}{Department of Physics and Astronomy, Johns Hopkins University, 3400 N. Charles Street, Baltimore, MD 21218}
\altaffiltext{3}{NASA Goddard Space Flight Center, Laboratory for High Energy Astrophysics, Code 662, Greenbelt, MD 20771}
\altaffiltext{4}{NASA Goddard Space Flight Center, Laboratory for Astronomy \& Solar Physics, Code 681, Greenbelt, MD 20771}

\begin{abstract}
We report the detection of radio emission coincident with the ultraluminous X-ray source (ULX) in Holmberg~II. The radio emission is diffuse and resolved, covering an area $\sim60~ \times 40$ pc in extent and well-matched to the recently discovered He{\scshape ~ii} nebula surrounding the X-ray source. Comparison of the radio and optical properties of this extended radio emission argue against its association with either an H{\scshape ~ii} region or supernova remnant. This is additional evidence that this ULX is not powered by a stellar-mass object whose emission is relativistically beamed towards the observer, and thus is either a super-Eddington source or intermediate mass black hole as suggested by optical observations. Implications of this result to future and existing radio studies of ULXs are discussed.
\end{abstract}


\section{Introduction}

One of the more interesting discoveries of X-ray observatories is that of unusually strong, compact non-nuclear sources in galaxies. These non-nuclear sources may have luminosities exceeding $2 \times 10^{39}$ \ergss, meaning that should they result from isotropic emission at the Eddington limit the implied masses are larger than those expected for end products of normal stellar evolution \citep[about 20 M$_\sun$;][]{fryer2001}. They might then be ``intermediate mass black holes'' \citep[IMBHs,][]{colbert1999}, an attractive possibility as these could fill in the black hole mass function between stellar-mass black holes and the supermassive ($10^6 - 10^9$ \msun) black holes found in the centers of galaxies. Similarly, there is a large gap in the distribution of X-ray luminosities for known discrete sources between X-ray binaries ($\lesssim10^{38}$ \ergss) and AGN ($\gtrsim10^{42}$ \ergss). Thus, ULXs are also often called intermediate X-ray objects or IXOs. It is clear that if ULXs proved to be associated with IMBHs in the $10^2 - 10^4$ \msun{} range, they would no longer violate the Eddington limit.

However, the identification of ULXs as IMBHs is only one possible theory. One obvious possibility is that ULXs are not physically associated with the host galaxies, being background AGN (or foreground stars) whose luminosity is thereby mistaken. While this may certainly explain some ULXs, the number of possible ULXs is well in excess of what would be expected based on random superpositions \citep{fabbiano1989}. Supernova remnants (SNR) are capable of producing X-ray luminosities of $\sim10^{40}$ \ergss, and are viable explanations for some ULXs \citep[e.g.,][]{fabian1996}. However, SNR do not exhibit the X-ray variability observed for most ULXs, which argues that ULXs are compact sources and thereby presumably XRBs.\footnote{Observed X-ray variability also largely precludes ULXs representing multiple unresolved XRBs, each of normal luminosity \citep{roberts2000}.} Thus, an attractive alternative are the more energetic hypernovae \citep[potentially associated with gamma-ray bursts;][]{fryer1998}, which could form both a compact X-ray object and a large associated optical nebula \citep{wang2002}. In fact, some ULXs have been demonstrated to be associated with large optical nebulae \citep[e.g.,][]{wang2002,pakull2003,roberts2003,kaaret2004a,lehmann2004}. In many cases, the X-ray spectra of ULXs are consistent with expectations for accreting black holes \citep[e.g., rotating Kerr holes,][]{makishima2000}. However, it is not required that these XRBs include a compact object more massive than a stellar mass black hole. \citet{king2001} proposed anisotropic emission from XRBs, thereby alleviating the need for super-Eddington emission. A similar model invokes relativistic beaming \citep{kording2002}, with ULXs akin to microquasars whose jets are observed directly along our line of sight. There is also a model which allows super-Eddington emission from thin accretion disks around stellar-mass black holes \citep{begelman2002}.

We are engaged in a study of archival NRAO Very Large Array (\vla~) data which aims to identify ULXs with radio counterparts. Radio emission is an excellent tool for understanding ULXs and differentiating among the various progenitor models. Interferometric radio data provide good positional accuracy, uncomplicated by the dust extinction which often plagues optical analyses. Under the assumption that the radio emission is powered by the same source as the X-ray emission, the radio morphology can determine whether the emission is beamed. Unresolved, compact radio emission is consistent with beaming towards the observer whereas extended emission rules out this possibility \citep[e.g., as in][]{kaaret2003}. Spectral indices garnered from multi-frequency observations may also be assessed relative to known sources and emission mechanisms (i.e., thermal emission or synchrotron). \citet{neff2003} noted the wealth of baseline expectations for the ratio of radio (at $\sim5$ GHz) to X-ray flux, enabling empirical comparisons between ULXs and more understood sources such as X-ray transients, SNR, star clusters, and AGN. Should some ULXs be associated with IMBHs, this last comparison is particularly apt. Perhaps most importantly, radio data enable the computation of source total energies and lifetimes. These may be assessed in relation to specific ULX progenitor models as well as the evolution of such sources. 

Among the better studied ULXs is source X-1 in the dwarf irregular galaxy Holmberg~II (or UGC~4305, DDO~50, and hereafter HoII). This ULX was first identified using the {\sl ROSAT} HRI \citep{zezas1999}, and has subsequently been observed using both \chandra{} \citep{kaaret2004a} and \xmm{} \citep{dewangan2004}. The proximity of HoII \citep[3.05 Mpc based on Cepheids, 3.39 Mpc from \hst{} observations of the red giant branch;][]{hoessel1998,karachentsev2002} has also made it a prime target for optical observation. \citet{pakull2002} noted He{\scshape ~ii} emission in the location of the ULX, which \citet{kaaret2004a} confirmed using \hst{} and \citet{lehmann2004} verified using several ground-based spectrographs. Both the X-ray data (in the spectral components and observed variation thereof) and the He{\scshape ~ii} data argue that the ULX in HoII is associated with a black hole with $L_x \gtrsim 4 \times 10^{39}$ \ergss, and therefore either a super-Eddington source or an IMBH of at least 25\msun{} \citep{dewangan2004,kaaret2004a}. In the present paper, we discuss the radio properties of the ULX in HoII. 

\section{Data}\label{sec-data}

HoII has been the subject of several \vla{} observing campaigns. Our objectives are to detect radio emission which is presumably faint, and do so at a resolution high enough to resolve modest physical scales \citep[1\arcsec{} = 14.8 pc at the distance of 3.05 Mpc assumed for HoII, chosen for consistency with][]{kaaret2004a}. Thus, we sought longer duration observations performed in the larger \vla{} configurations. Two general programs fit these parameters, one at 1.4~GHz in the A configuration and the other at 4.86~GHz in the B configuration (each with a resolution of $\sim$1\farcs5). There were also prior radio observations at $\sim$15\arcsec{} resolution at 335~MHz, 1.4~GHz, and 4.86~GHz detailed in \citet{tongue1995}. These detected unresolved radio emission consistent with the ULX position, with corresponding flux densities of 6.4$\pm$2.1 mJy, 1.37$\pm$0.08 mJy, and 0.77$\pm$0.05 mJy, respectively.

The 1.4~GHz data were collected on 3 April 1994 as program code AT159, and consist of over 3 hours time on source. The raw data were obtained from the archive and loaded including antenna-specific weights. The flux scale was set using 3C286 with phase calibration provided by 0903+679 (J2000), following recommendations in the {\it VLA Calibration Manual}.\footnote{{\url http://www.aoc.nrao.edu/$\sim$gtaylor/calib.html}} Imaging was done using the task {\scshape imagr} with multiple facets to alleviate problems such as the ``3D effect.'' The ({\it u,v}) data were gridded and tapered in a manner which resulted in a beam with good Gaussian characteristics, helping produce excellent sensitivity output maps. Multiple rounds of imaging and self calibration were performed until the final map was created. This map has an rms noise of just 17 $\mu$Jy beam$^{-1}$ in the vicinity of the ULX position, where the beam size is 1\farcs9$\times$1\farcs5.

\begin{figure}
\includegraphics[angle=270,scale=0.375]{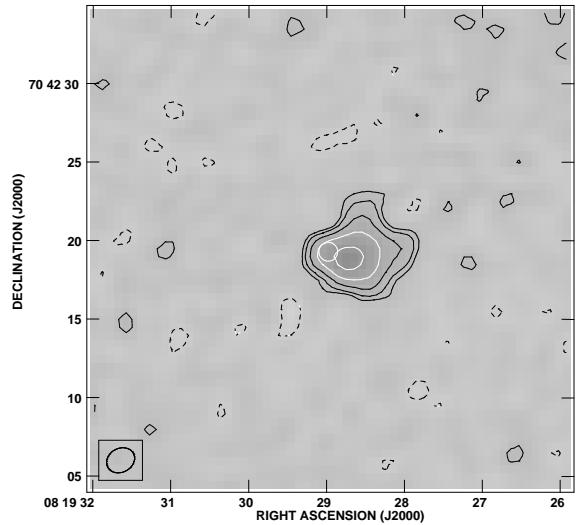}
\caption{1.4~GHz map, with contours plotted at -2, 2, 3, 5, 8, and 13 times a base level of 17 $\mu$Jy beam$^{-1}$. The restoring beam is shown at lower left. The \chandra{} ULX position is shown as a circle whose size represents its positional uncertainty.\label{fig-Lmap}}
\end{figure}

Radio emission coincident with the ULX location is readily apparent (see Figure \ref{fig-Lmap}). Based on a Gaussian fit (using the task {\scshape jmfit}), its peak is located at 08$^{\mbox{{\scriptsize h}}}$19$^{\mbox{{\scriptsize m}}}$28\fs65 $+$70\degree42\arcmin19\farcs0, about 1\farcs6 from the \chandra{} ULX position reported in \citet{kaaret2004a}. The error in the radio position is less than 0\farcs1, although the radio emission is clearly extended and covers the \chandra{} position. Based on the surface density of detected radio sources within the central 10\arcmin{} of the radio image, the probability of a chance ULX-radio superposition within 1\farcs6 is well under 0.1$\%$. The radio emission has deconvolved major and minor axes of 3\farcs7 (1$\sigma$ range of 3\farcs4 - 4\farcs0) and 2\farcs7 (2\farcs4 - 2\farcs9) at a position angle of 91\degree{} (80\degree{} - 104\degree). The flux is 1.174$\pm$0.085 mJy, after correction for primary beam attenuation. Emission on slightly larger scales is possible given the 1.4~GHz flux reported in \citet{tongue1995}, although the two fluxes do not differ significantly.

The 4.86~GHz data consist of three related programs totalling four observations. Unfortunately, each observation is relatively short at $\sim$10 minutes time on source. Program AR165 observed HoII on 18 January 1988 and 14 March 1989, while program AR227 and AR258 observed it on 29 July 1990 and 6 December 1991, respectively. Each observation was performed with effectively identical observational setups, making it fairly easy to combine the data into a single $\sim40$ minute observation as well as analyzing each individually.

\begin{figure}
\includegraphics[angle=270,scale=0.375]{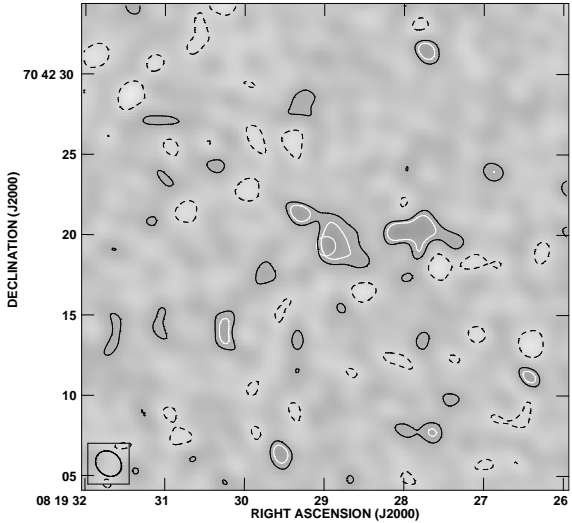}
\caption{4.86~GHz map. All contours and symbols the same as Figure \ref{fig-Lmap}, with the exception of the base contour level being 44 $\mu$Jy beam$^{-1}$.\label{fig-Cmap}}
\end{figure}

Data reduction was achieved using the same basic procedures as the 1.4~GHz data except in this case the phase calibrator was 0642+679 and the lack of strong sources within the fields prevented self-calibration. The ULX was not significantly detected in any of the individual observations, with rms noises in the ULX vicinity ranging from 58 to 79 $\mu$Jy beam$^{-1}$. Peaks in the radio emission within 2\arcsec{} of the ULX \chandra{} position were determined, and although the flux between the different observations differed by a factor of two this variability is not statistically significant. At this point, the ({\it u,v}) data for the four observations were combined, weighting the individual observations by the inverse of their rms noise squared. The rms noise of the final map created using the combined data was 44 $\mu$Jy beam$^{-1}$ with a beam size of 1\farcs7$\times$1\farcs5. Weak extended emission having an integral flux of 0.677$\pm$0.207 mJy covers the ULX position (see Figure \ref{fig-Cmap}). The deconvolved size is 3\farcs4 (2\farcs3 - 4\farcs5) by 1\farcs9 (1\farcs0 - 2\farcs) with a position angle of 37\degree (10\degree{} - 66\degree). The peak in this emission, located at 08$^{\mbox{{\scriptsize h}}}$19$^{\mbox{{\scriptsize m}}}$28\fs84 $+$70\degree42\arcmin19\farcs5 (accurate to $\sim$0\farcs2 and within 0\farcs7 of the \chandra{} position), does not constitute a traditional 5$\sigma$ detection. However, imaging runs using pure natural weighting of the ({\it u,v}) data with a slight taper were able to yield detections at over 6$\sigma$ significance. The integral flux of the source was consistent in all imaging runs and additionally is consistent with the value reported in \citet{tongue1995}.

Taken together, the 1.4~GHz and 4.86~GHz data imply a spectral index of $\alpha = 0.44 \pm 0.31$ (where $S_\nu \propto \nu^{-\alpha}$). \citet{tongue1995} determined a spectral index of 0.5$\pm$0.1 over the same frequency range for the region coincident with the ULX, based on their lower resolution data. The greater uncertainty in our value is simply the result of the fairly noisy 4.86~GHz data.

\section{Discussion}\label{sec-discuss}

\begin{figure}
\includegraphics[angle=270,scale=0.375]{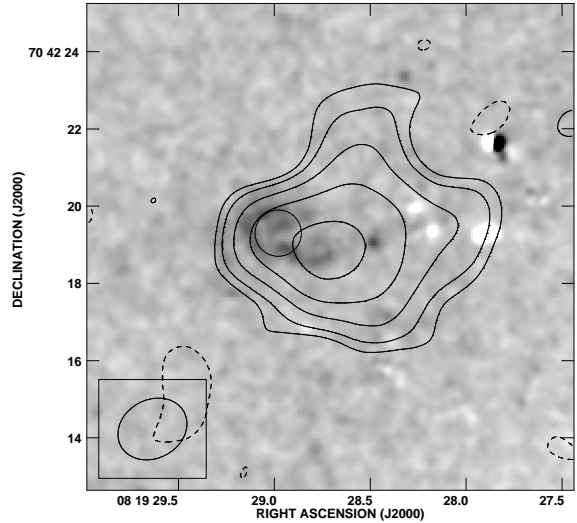}
\caption{1.4~GHz contours overlaid on He{\scshape ~ii} emission as presented in \citet{kaaret2004a}. The \chandra{} position is indicated by a circle sized to represent the positional uncertainty. The H{\scshape ~ii} region extends well to the west of this figure.\label{fig-radhe}}
\end{figure}

The location of the ULX in HoII has been shown to lie within extended radio emission. Should the radio emission arise from the same physical source as the X-ray emission, the resolved nature of the radio argues against relativistic beaming along the line of sight. In fact, the size of the radio emission implies a region of $\sim60$pc by $\sim$40pc, elongated in the East-West direction. The eastern region of the radio emission is coincident with the He{\scshape ~ii} nebulae reported in \citet{kaaret2004a}, as shown in Figure \ref{fig-radhe}.

The obvious question here is to what extent the radio emission is linked to the ULX. The ULX position is associated with a region of star formation indicated by H$\alpha$ and other emission lines \citep[e.g.,][]{puche1992,kaaret2004a,lehmann2004}, and H{\scshape ~ii} regions are known sources of radio emission through thermal bremsstrahlung and non-thermal emission arising in supernovae. The derived radio spectral index is marginally steeper than that of bremsstrahlung which has $\alpha \approx 0.1$, suggesting that all the emission is not simply due to this. We can also get some idea of the contribution of bremsstrahlung to the net radio flux through the reported emission line measurements. \citet{kaaret2004a} find a net flux of $1.1 \times 10^{-14}$ \ergss{} for H$\beta$, and \citet{lehmann2004} find that the H{\scshape ~ii} region has little extinction (i.e., the ratio of H$\alpha$ to H$\beta$ is close to 2.85). For thermal bremsstrahlung emission in H{\scshape ~ii} regions, \citet{moorwood1983} derive the expression:
\begin{equation}
I_{Br~\alpha} = 2.71 \times 10^{-14} ( \frac{T}{10^4} )^{-0.85} ( \frac{\nu}{5~GHz} )^{0.1} F_\nu
\end{equation}
Using a line intensity ratio of 0.0834 for $j_{~\mbox{Br}\alpha} / j_{~\mbox{H}\beta}$ (e.g., Osterbrock for $T = 10^4$ K), this predicts a 4.86~GHz flux of only 0.034 mJy. Thus, the likely contribution of thermal bremsstrahlung to the total radio emission is quite small, about $5\%$. Qualitatively, we also note that the radio and He{\scshape ~ii} emission appear only at the eastern edge of the full H{\scshape ~ii} region, which is $\sim15$\arcsec{} in size. Furthermore, the He{\scshape ~ii} emission line is narrow and lacks the associated ``bump'' from nearby N{\scshape ~iii}, C{\scshape ~iii}, and C{\scshape ~iv} seen in Wolf-Rayet spectra. That the line is narrow also shows the He{\scshape ~ii} is optically thin, further underscoring that the calculated lower bound on the X-ray luminosity ($4 \times 10^{39}$ \ergss) of the source reported in \citet{kaaret2004a} is indeed a true lower limit.

Alternatively, we may consider SNR. The total 4.86~GHz flux is remarkably comparable to what would be emitted by Cas~A were it to lie within HoII, and the derived spectral index is consistent with shell-type supernovae \citep[$\alpha=0.45$,][]{clark1976}. However, the radio morphology does not exhibit a shell-type appearance but rather is filled, or ``plerionic'' \citep[e.g.,][]{weilerpan1978}. These types of SNR have flatter spectral indices ($\alpha\sim0$) and are of much smaller size. In fact, the radio-emitting region around the HoII ULX is very large for its luminosity, placing it as an outlier to any $\Sigma - D$ relation \citep[e.g.,][]{green2004}. Similarly, the integral field spectroscopy of \citet{lehmann2004} indicates that the line emission is not that expected for supernovae nor is the X-ray emission consistent with an SNR origin \citep[e.g.,][]{dewangan2004}.

The radio data also allow for calculation of total energy and cooling time under the assumption of equipartition \citep[e.g.,][]{longair}. For this calculation, we made the simple approximation that the radio source was a sphere whose projected area is equivalent to the measured size at 1.4~GHz, and that the filling factor within this sphere was unity (the effect of lowering the filling factor is to reduce the minimum energy and cooling time). The total radio luminosity of the source was estimated by assuming a power law between the observed frequencies of 1.4~GHz and 4.86~GHz. The resulting minimum energy was $2.6 \times 10^{49} \eta^{4/7}$ ergs, implying a cooling time of $2.5 \times 10^7 \eta^{4/7}$ years (where $\eta$ is the ratio of baryon to electron particle energy). The calculated minimum magnetic field strength was $13\eta^{2/7}$ $\mu$G. Derived cooling times under the assumption that the radio emission is thermal bremsstrahlung are about an order of magnitude larger. 

The collection of X-ray, optical, and radio data show that the radio emission is not due to a simple SNR or H{\scshape ~ii} region, and that the radio emission is positionally associated with the X-ray source and the He{\scshape ~ii} nebula. We are thus led to the conclusion that the radio emission and the ULX are causally connected. In searching for other examples of accreting binaries with extended luminous radio emission, we note some similarities to the Galactic system of the W50 nebula and the jet source SS~433. In each case, the radio emission covers an extent of $\sim50$ pc with a total spectral index of about 0.5 \citep[e.g.,][]{dubner1998}. However, the HoII nebula is over an order of magnitude more luminous and correspondingly implies a much greater energy in relativistic particles \citep[$5 \times 10^{45} \eta^{4/7}$ ergs for W50-SS~433 vs. $2.6 \times 10^{49} \eta^{4/7}$ ergs for HoII;][]{dubner1998}.

Computation of a cooling time also enables an investigation of possible source motion as discussed in \citet{kaaret2004b}. The simple hypothesis is that ULXs are associated with stellar clusters, and their X-ray emission declines as they move away from these clusters. The HoII ULX is bright, but not close to any apparent stellar cluster. Thus, for this model to be correct the HoII ULX would need to be travelling at high velocity. Using the major axis size of $\sim60$ pc and the cooling time of $2.5 \times 10^7$ years, the velocity is only about 2 km s$^{-1}$. This is much less than the expected kick velocities of X-ray binaries (unless there is a very substantial radial component to the velocity), and would seem to preclude the ULX being a young system ejected from a stellar cluster at high velocity.

What are the prospects for detecting similar ULXs in the radio? Based on the X-ray luminosites reported in \citet{kaaret2004a} and our radio results, a representative value of $R_X$ \citep[the radio to X-ray ratio, see][]{terashima2003,neff2003} for the HoII ULX is $5 \times 10^{-6}$ \citep[about the same as for the NGC~5408 ULX][]{kaaret2003,neff2003}. Assuming this $R_X$ and $L_X = 10^{39}$ \ergss{}, a one-hour integration with the \vla{} can detect such a source at the 5$\sigma$ level provided it lies within a distance of 2.8 Mpc {\it and is unresolved}. If this hypothetical source has an intrinsic size of 40 pc (hence about 3\arcsec{} across), the one-hour observation will fail to detect it unless the \vla{} is in its more compact C or D configurations. 

The above explanation fits the results of two other radio studies of ULXs. \citet{kaaret2003} used archival ATCA data to detect the ULX in NGC~5408 at 4.8~GHz (where it was unresolved) but not at 8.64~GHz. The resolutions were 2\farcs9 $\times$ 1\farcs8 and 1\farcs6 $\times$ 1\farcs0, respectively \citep[see][]{stevens2002}. At the assumed distance of 4.8 Mpc, $1\arcsec{} = 23$ pc. Given the noted 40 $\mu$Jy beam$^{-1}$ rms of the 8.64~GHz data, emission extended over $\sim$40 pc might easily be missed. This scenario is roughly consistent with the spectral index of the HoII data, provided the 4.8~GHz NGC~5408 emission is only marginally unresolved. This theory will be testable in the near future, as an NGC~5408 monitoring campaign using the \vla{} is currently in progress (Kaaret, private communication). Similarly, \citet{kording2005} performed a monitoring campaign of nine nearby ULXs (within 5.5 Mpc) with possible detections of two compact sources, believed to be SNR. The observations were performed with the \vla{} primarily in its largest configuration (A) at 8.46~GHz, with a resolution of about 0\farcs25. Their excellent sensitivities of $\sim15$ $\mu$Jy beam$^{-1}$ are still nearly an order of magnitude larger than necessary to detect any diffuse emission akin to that observed for HoII.

In summary, we have identified an extended radio source coincident with the ULX in HoII. The peak of the radio emission lies within 1\farcs6 of the ULX location, implying a very small probability that the radio and X-ray sources represent a random superposition. The properties of the radio emission argue against simple interpretations as an H{\scshape ~ii} region or SNR, particularly when considered in light of the He{\scshape ~ii} nebula and attendant optical spectroscopy. We believe that this association, very different from what has been observed for accreting X-ray sources within the Galaxy, furthers the identification of HoII X-1 as an intermediate mass black hole.

\acknowledgements
The authors thank Michael Rupen for supplying the 4.86~GHz radio data, Schuyler Van Dyk for a preliminary 1.4~GHz image, Philip Kaaret for the He{\scshape ~ii} image, and an anonymous referee for insightful comments.

\end{document}